\documentclass[12pt]{iopart}

\usepackage{iopams}

\newcommand{\x}{\mathsf{X}}

\begin{document}

\title[Auxiliary fields ...]{Auxiliary fields in the geometrical 
relativistic particle dynamics.}

\author{A Amador\dag \, 
N Bagatella\ddag \, 
R Cordero\P \,and E Rojas\S}

\address{\dag\ddag\S Departamento de F\'\i sica, Facultad 
de F\'\i sica e Inteligencia Artificial, Universidad 
Veracruzana, 91000 Xalapa, Veracruz, MEXICO}
\address{\P Departamento de F\'\i sica, Escuela Superior de 
F\'\i sica y Matem\'aticas del I.P.N., Edificio 9, 07738 M\'exico 
D.F., MEXICO}

\eads{\mailto{aramador@gmail.com},\,
\mailto{nbagatella@uv.mx},\,
\mailto{cordero@esfm.ipn.mx},\, \mailto{efrojas@uv.mx}}

\begin{abstract}
We describe how to construct the dynamics of relativistic 
particles following, either timelike or null curves, 
by means of an auxiliary variables method instead of the 
standard theory of deformations for curves. There are 
interesting physical particle models governed by 
actions that involve higher order derivatives of the 
embedding functions of the worldline. We point out that the 
mechanical content of such models can be extracted wisely
from a lower order action, which can be performed by 
implementing in the action a finite number of constraints 
that involve the geometrical relationship structures 
inherent to a curve and by using a covariant formalism. We 
emphasize our approach for null curves. For such systems, 
the natural time parameter is a pseudo-arclength whose properties 
resemble those of the standard proper time. We illustrate the 
formalism by applying it to some models for relativistic particles.
\end{abstract}

%\submitto{\JPA}

\pacs{14.80.-j, 02.40.Hw, 04.50.-h}

%\maketitle

\date{\today}

\section{Introduction}

The first geometrical models for rigid particles result as a 
byproduct of the point-like versions for highly dimensional 
models that involve the extrinsic curvature of the worldvolume 
swept out by relativistic strings or branes
%\footnote{The term
%rigidity refers to the dependence in the geodesic curvature or 
%extrinsic curvature of the Lagrangians that describe the dynamics 
%of this sort of particles or branes.}  
\cite{Polyakov}. 
Thenceforth, the interest in this sort of particle models have 
grown by leaps and bounds because one can find potential applications 
both in particle physics and mathematics. For instance, they can 
describe spinning particles, whether massive or massless, defined 
on timelike trajectories \cite{Plyuschay,Nes}, and when the model 
is linear in the geodesic curvature it turns out to be related to 
a massless particle with $W_3$ gauge symmetry \cite{ramos}. 
The story did not end there. Recently, Nersessian and Ramos 
proposed certain models for massive particles associated with null 
curves \cite{nerse1,nerse2}. Immediately, considerable  effort has 
been devoted by other authors in the understanding of the geometry 
of these models as well as its applications \cite{ferra1,ferra2,ferras}. 
However, a key drawback for all these models resides in their higher order 
derivative nature, as a consequence physicists have been reluctant 
to consider their study due to technical difficulties like the 
increasing of the degrees of freedom and the equations of motion 
being at least of fourth-order in derivatives of the fields which 
do not appear to be tractable. Certainly this 
unpleasant fact appears to be a great difficulty but these models 
have the advantage of encode the spin content of the particles in 
the geometry of the worldlines. The standard way to studying this 
sort of particles is through the theory of deformations sheltered 
by a Frenet-Serret (FS) basis adapted to the worldline. 
Unfortunately this gives raise to lengthy and annoying computations 
due to the above mentioned non-trivial higher order derivative property 
inherent to rigid particles \cite{ACG,nesterenko}. In this paper 
we aim to study a powerful tool for the worldline geometry either 
timelike or lightlike, namely, the conserved linear momentum whose 
existence is a simple consequence of the Noether theorem. 
A striking property of the stress tensor for particles or extended 
objects is that its conservation in time yields not only the equations of 
motion but also the intrinsic geometrical properties for every model 
under consideration \cite{Noether}. To overcome the majority of the 
typical technical obstacles in the obtaining of the rigid particle 
dynamics, we appeal to an auxiliary variables method that was originally
introduced for the study of general surfaces and applied to describe 
fluid membranes \cite{jemal}. Even though most of the progress in 
the study of particle models has been made in the spirit of the 
standard theory of deformations, the conserved linear momentum has not 
been exploited completely in this context. Therefore, we provide an 
alternative way to analyse point particle models by means of an easy 
obtaining of the conserved linear momentum. The main idea behind the 
work is to replace the original action by one equivalent depending 
on lower order derivatives evading in this way the standard theory 
of deformations. We assure that this approach simplifies the dynamical 
point particle description. 

The outline of the paper is as follows. In Sect. 2 we begin with a 
glimpse of the worldline Frenet-Serret geometry describing both, 
timelike and lightlike, particle trajectories. This brief 
section will serve mainly to explain our notation and the basic facts 
to be used in this paper. In Sect. 3 we apply an auxiliary variables 
method to obtain the conserved linear momentum associated to a local 
geometrical action depending of the geodesic curvature and the torsion.
We emphasize our approach for the case of null curves since the 
existing point particle models with this geometry are not widely known.
We conclude in Sect. 4 by mentioning some comments. We have tried 
throughout the paper to follow an index-free notation in order to avoid a
cumbersome notation. Definitions of constructed deformations which are 
helpful to understand the geometrical nature of a particle worldline 
and important identities of the theory of deformations for curves have 
been collected in Appendix A. To complement our approach in the null 
case we obtain the Casimir invariants associated to the Poincar\'e 
symmetry which is the subject of Appendix B. In our context, these are 
useful to integrate the equations of motion.

\section{Worldline geometry}

\subsection{Timelike curves}

Consider a relativistic particle whose timelike worldline can be 
described by the embedding  $x^\mu = X^\mu (\xi)$, where $x^\mu$ 
are local coordinates in Minkowski spacetime with metric $\eta_{\mu \nu}= {\mbox{diag}}(-1,+1,+1, \ldots ,+1)$ and $(\mu ,\nu = 0,1, \ldots,N-1)$, 
$\xi$ is an arbitrary parameter and $X^\mu$ are the embedding functions. 
The vector tangent to the worldline is given by $\dot{X}^\mu = dX^\mu/d\xi$ 
such that the one-dimensional metric along the curve is $\gamma = 
\eta_{\mu \nu}\dot{X}^\mu \dot{X}^\nu \equiv \dot{X} \cdot \dot{X}$. We 
assume that for timelike curves $\dot{X}^2 < 0$ is satisfied. The 
infinitesimal arclength for the worldline is given by 
\begin{equation}
d\tau = (-\dot{X}\cdot \dot{X})^{1/2}d\xi\,. 
\label{eq:1}
\end{equation}
This arclength is invariant under reparametrizations of the worldline. 
We introduce $N-1$ normal vectors to the worldline, denoted by 
$n^\mu {}_i \quad (i=1,2,\ldots,N-1)$. These are defined implicitly by 
$n^i \cdot \dot{X} = 0$ and normalized as $n_i \cdot n_j = \delta_{ij}$. 

Though we may choose to label points along the curve arbitrarily, the most 
convenient approach to study the dynamics for relativistic particles is to let 
the parameter to be the arclength along the worldline. We will denote with a 
prime differentiation with respect to $\tau$. Therefore we introduce the 
orthonormal basis $\left\lbrace X', \eta_i \right\rbrace $ which satisfy 
$X'\cdot X' = -1, \,\,\,X'\cdot \eta_i = 0$ and $\eta_i \cdot \eta_j = 
\delta_{ij}$. This basis obeys the following $N$-dimensional FS equations 
\cite{ACG,doCarmo}
\begin{eqnarray}
 X'' &=& k_1 \eta_1 \,, \nonumber  \\
\eta_1 ' &=& k_1 X' - k_2 \eta_2 \,, \nonumber \\
\eta_2 '&=& k_2 \eta_1 - k_3 \eta_3\,, \nonumber \\
\ldots && \ldots 
\label{eq:FS}
\\
\eta_{N-2} ' &=& k_{N-2} \eta_{N-3} - k_{N-1} \eta_{N-1}\,,
\nonumber \\
\eta_{N-1} ' &=& k_{N-1} \eta_{N-2}\,,\nonumber 
\end{eqnarray}
where $k_i$ stands for the independent $i$th FS curvature and $k :=k_1$ is 
known as the {\it geodesic curvature}. 
Note that from the FS equations (\ref{eq:FS}) we can express the 
geodesic curvature as
\begin{equation}
k_1 = - X' \cdot \eta_1 '\,.
\label{eq:k}
\end{equation}
Also note that the geodesic curvature is given in terms of second order derivatives
of the embedding functions, $k_1 = \sqrt{X''\cdot X''}$.

\subsection{Lightlike curves}

We turn now to consider a null curve, for the sake of simplicity, 
in a $3+1$ ambient Minkowski spacetime with metric $\eta_{\mu \nu}$ 
described by the embedding $x^\mu = \x^\mu (\rho)$ where $x^\mu $ 
are local coordinates in the background spacetime, $\rho$ is an 
arbitrary parameter and $\x^\mu$ are the embedding functions $(\mu 
= 0,1,2,3)$. Hereafter, in order to compare with respect to the 
timelike case (see for instance, (\ref{eq:1}) and (\ref{eq:pseudo})), 
we consider the signature of $\eta_{\mu \nu}$ to be $(+,-,-,-)$. With 
this convention timelike vectors have a positive norm. The tangent 
vector to the curve is given by $ \dot{\x}^\mu = d\x^\mu /d\rho$. It 
satisfies that $\dot{\x}\cdot\dot{\x}=0$ since the curve lies on the 
light cone so the arclenght vanishes. This null condition 
on the tangent vectors shatters our accustomed vision of the 
worldline geometry which leads us to promote $\Upsilon = 
\ddot{\x}\cdot \ddot{\x}$ as the corresponding worldline metric 
\cite{nerse1}. This new point of view necessarily forces
the introduction of a new parameter called pseudo-arclength which becomes 
fruitful to normalize the derivative of the lightlike tangent 
vector \cite{nerse1,null}. 
The infinitesimal pseudo-arclength for a null curve is given by
\begin{equation}
 d\sigma = (-\ddot{\x}\cdot\ddot{\x})^{1/4}d\rho\,.
\label{eq:pseudo}
\end{equation}
This pseudo-arclength is invariant under reparametrizations of the curve. 
We shall use again a prime to denote derivation with respect to $\sigma$. 
To analyse the geometry of null curves is desirable to adapt a FS frame 
constructed in a similar way as in the timelike case \cite{nerse1,ferra1,null}. 
In such approach we consider a basis adapted to null curves spanned 
by $\left\lbrace e_+,e_1,e_-,e_2 \right\rbrace $ where $e_+$ and $e_-$ are 
lightlike whilst $e_1$ and $e_2$ are spacelike. The null FS basis has 
the structure
\begin{eqnarray*}
e_+ & = \x'\,,\\
e_+ ^2 &= e_- ^2 = 0\,,\\
e_\pm \cdot e_1 &= e_\pm \cdot e_2 = e_1 \cdot e_2 =0\,, \\
e_+ \cdot e_- &= -e_1 \cdot e_1 = - e_2 \cdot e_2 = 1\,.
\end{eqnarray*}
This basis obey the following 4-dimensional FS equations 
\cite{nerse1,ferra1,null}
\numparts
\begin{eqnarray}
 {e}_+ ' &=& e_1\,,\\
{e}_1 ' &=& \kappa_1\,e_+ + e_-\,,
\label{eq:NFS2}
\\
{e}_- '&=& \kappa_1 \,e_1 + \kappa_2 \,e_2\,,
\label{eq:NFS3}\\
{e}_2 ' &=& \kappa_2 \,e_+ \,,
\end{eqnarray}
\endnumparts
where $\kappa_1$ and $\kappa_2$ are independent curvature functions of the 
null curve similarly as in the timelike case. Occasionally, $\kappa_1$
is known as the torsion due to its dependence of the third-order 
derivatives of the field variables. Note that the torsion can be 
expressed in several forms. For our purposes below, one convenient 
way is 
\begin{equation}
\kappa_1 = \frac{1}{2} e_+ '' \cdot e_+ ''\,. 
\label{eq:kappa1}
\end{equation}
It is worth noting that $\kappa_1$ is given in terms of the 
third-order derivatives of the field variables, $2\kappa_1 =  
(\x''' \cdot \x''')$.

\section{FS dynamics}

\subsection{Timelike case}

We assume that the dynamics of a rigid particle is specified by an action 
invariant under reparametrizations of the timelike worldline of the form
\begin{equation}
 S_0 [X] = \int d\tau \,L(k_1)\,,
\label{eq:action}
\end{equation}
where $L$ is a scalar under reparametrizations. It is usual that
under an infinitesimal deformation of the embedding $X \to X + 
\delta X$, the response of the functional (\ref{eq:action}) 
casts out the equations of motion and the Noether charges 
\cite{ACG,Noether}. To accommodate an auxiliary variables method 
describing rigid particles we follow the seminal work given in 
\cite{jemal}. We would like to distribute this deformation among 
the parametrization, the FS basis and $k_1$. That is why we consider 
them as new independent variables. To promote them as intermediate 
auxiliary variables it is necessary to implement constraints 
involving their definitions smeared out with Lagrange multipliers. 

Thus, we construct now a new functional 
action $S [k_1,\eta_1, X', X,f,\lambda^1,\lambda^{11},\lambda]$ of 
the form
\begin{eqnarray}
\fl 
S= S_0[X,k_1] + \int d\tau f\cdot \left( X' -\frac{dX}{d\tau}\right) 
+ \int d\tau \,\left[ \lambda^1 \,\left(X' \cdot \eta_1 \right) +
\lambda^{11}\, \left(\eta_1 \cdot \eta_1 - 1 \right) 
\right] \nonumber \\ 
\lo + \int d\tau \, \lambda\, \left( k_1 + X' \cdot \eta_1 ' \right)  \,.
\label{eq:Caction}
\end{eqnarray}
This is a suitable departure point which provides both 
geometrical and physical insights into the mechanical 
systems described by (\ref{eq:action}) by means of a conserved 
linear momentum. 

The Euler-Lagrange (EL) derivative for $X$ is such that in the 
extremum condition it shows a conservation law,
\begin{equation}
 \frac{d}{d\tau}\left\lbrace  f^\mu + \left[ L + \lambda \,k_1
+ (f \cdot X') \right]\,X^{'\,\mu}  \right\rbrace = 0\,,
\label{eq:conserv}
\end{equation}
where we have employed the identities (\ref{eq:T1}) and the expression
(\ref{eq:k}).

The EL derivative associated to $X'$ exhibits the 
geometrical form of $f^\mu$ in terms of the Lagrange multipliers
and the FS basis
\begin{equation}
 f = -\lambda\, k_1\,{X'}
- \lambda^1 \,\eta_1 + \lambda\,k_2\,\eta_2 \,,
\label{eq:f1}
\end{equation}
where we have used the expressions (\ref{eq:FS}). As a result we
obtain $(f\cdot X')=\lambda\,k_1$. Correspondingly, the EL 
derivative for $\eta_1$ and by exploiting the FS equations 
(\ref{eq:FS}) we have
\begin{equation}
\eqalign{
 \lambda^1 = \lambda'\, ,\\
2 \lambda^{11} = \lambda\,k_1\,.
}
\label{eq:Lm}
\end{equation}
Finally, the EL derivative for $k_1$ yields
\begin{equation}
\lambda = - L^*
\end{equation}
where we have introduced the notation $L^* = d L/d k_1$. Thus, 
we can identify the Lagrange multipliers (\ref{eq:Lm}) as
$\lambda^1 = - L^{*\,'}$ and $2\lambda^{11} = - k_1\,L^*$.

Hence, putting all of these results together in the conservation law 
(\ref{eq:conserv}) we therefore get
\begin{equation}
 \frac{d}{d\tau} \left[ (L - L^* k_1)\,X' + L^{*\,'}\,
\eta_1 - L^* k_2\,\eta_2 \right] =0\,,
\label{eq:eom}
\end{equation}
which allows us to identify the conserved linear momentum, 
written in terms of the FS basis,
\begin{equation}
 p = (L - L^* k_1)\,X' + L^{*\,'}\,
\eta_1 - L^* k_2\,\eta_2 \,.
\label{eq:fmu1}
\end{equation}
This is nothing but the linear momentum associated to the
Noether charge specialized to a constant infinitesimal translation
$\delta X^\mu = \epsilon^\mu$, \cite{ACG}. Further, the momentum 
(\ref{eq:fmu1}) is in accordance with the momentum conjugated to the 
embedding variables in an Ostrogradski Hamiltonian approach for the 
action (\ref{eq:action}) \cite{hamFS}. 

The FS projections of the total derivative (\ref{eq:eom}) permit to 
deduce the mechanical and geometrical properties of the generic action 
(\ref{eq:action}). The projection of (\ref{eq:eom}) along $\eta_3$ 
implies the vanishing of $k_3$ thereby the motion is performed in 
$2+1$ dimensions. Similarly, the projection of (\ref{eq:eom}) along 
$\eta_2$ leads to $(L^*)^2k_2 = $const., which can be interpreted as a 
conservation of the spin of the particle \cite{ACG}. To finish the 
tangential projection casts out the equations of motion, namely, 
$L^{*\,''} + (L - L^*\,k_1)\,k_1 - L^* \,k_2 ^2 =0$. These properties 
as well as the Poincar\'e invariants have been well discussed in 
\cite{ACG,nesterenko}.
 
In closing this subsection, we apply the formalism developed above 
to the linear correction to the free relativistic particle, $L = 
-m + \alpha k_1$, where $m$ and $\alpha$ are constants. Obviously 
we have $L^* = \alpha$. The corresponding linear momentum is given by 
$p = - m \,X ' -\alpha k_2\, \eta_{2} $ and the equation of motion results 
$\alpha m k_1 + {\mbox{{\small const.}}}=0$.

\subsection{Lightlike case}

Now, we shall consider actions for null curves that are invariant under 
reparametrizations of the form
\begin{equation}
S_0 [\x]= \int d\sigma \,L(\kappa_1)\,,
\label{eq:null-act}
\end{equation}
where $L$ is invariant under worldline reparametrizations.
An auxiliary variables method will distribute the deformation 
$\x \to \x + \delta \x$ among $\x$ itself, $e_+$ and $\kappa_1$ 
considering all of them as new independent variables. Once 
again, bearing in mind the necessity of promote them as 
auxiliary variables we need to implement them
through their definitions and structure properties 
smeared with appropriated Lagrange multipliers.

Following the timelike case, we now construct the functional 
$S[\kappa_1,e_+, \x , \mathsf{f},\Lambda_{++},\Lambda ]$ written as
\begin{equation}
\fl
S= S_0[\x,\kappa_1] + \int d\sigma \,\,\mathsf{f}\cdot \left( e_+ - 
\frac{d}{d\sigma}\x\right) + \int d\sigma \,\,\Lambda_{++}\,e_+ ^2 
+ \int d\sigma \Lambda \,\left( \kappa_1 - \frac{1}{2} e_+ '' \cdot e_+ ''
\right).
\end{equation}

A direct computation of the EL derivative for $\x$ shows that in the 
extremum condition we have 
\begin{equation}
 \frac{d}{d\sigma}\left\lbrace \mathsf f^\mu - \frac{1}{2} 
\frac{d}{d\sigma}  \left[ \left( L + 4\Lambda\,\kappa_1 + 
(\mathsf f\cdot e_+) \right) e_1^\mu \right]   \right\rbrace  = 0\,,
\label{eq:law}
\end{equation}
where we have employed the identities 
(\ref{eq:TT2}) and the expression (\ref{eq:kappa1}).
The EL derivative for $e_+$ allows us to write $\mathsf f^\mu$ in terms
of the null FS frame as
\begin{equation}
\mathsf f =  \left( \Lambda\,e_1 '\right)''  - 2\,\Lambda_{++}\,e_+ \,. 
\end{equation}
By making use of the null FS equations we obtain
$(\mathsf f \cdot e_+)= 2\Lambda\,\kappa_1 + \Lambda''$.
Finally, we compute the EL derivative with respect to $\kappa_1$,
\begin{equation}
 \Lambda = - L^*\,,
\end{equation}
where we have used one more time $*$ to denote derivative with respect to 
$\kappa_1$, i.e., $L^* = dL/d \kappa_1$. We are ready to insert the information 
into the conservation law (\ref{eq:law}). With the previous results we obtain 
$\mathsf f = - \left(L^*\,e_1 '\right) '' - 2\Lambda_{++}\,e_+ $, such that 
the conservation law (\ref{eq:law}) reads
\begin{equation}
\mathsf E= \frac{d}{d\sigma}\left\lbrace  2\Lambda_{++}\, e_+
+ (L^*\,e_1 ')'' + \frac{1}{2} \frac{d}{d\sigma} 
  \left[ \left( L - 6 L^*
\kappa_1 - L^{* ''} \right)  e_1 \right] \right\rbrace  = 0\,,
\label{eq:eom-n}
\end{equation}
which helps to identify the corresponding linear momentum, 
$\mathsf p = \mathsf p _+ e_+ + \mathsf p _- e_- + 
\mathsf p _1 e_1 + \mathsf p _2 e_2$, in the null FS frame. 
If the conservation law (\ref{eq:eom-n}) is expressed as $\mathsf E^\mu = 
\mathsf p^{\mu\,'}=0$, it is straightforward to obtain the conditions that the 
momentum components must satisfy,
\begin{equation}
\eqalign
{
\mathsf p _+ ' + \mathsf p _1 \kappa_1 + \mathsf p _2 \kappa_2 =0\,,\\
\mathsf p _- \kappa_2 + \mathsf p _2 ' = 0\,,
}
\label{eq:condi1}
\end{equation}
and 
\begin{equation}
\eqalign
{
\mathsf p _1 + \mathsf p _- ' = 0\,,\\
\mathsf p _+ + \mathsf p _1 ' + \mathsf p _-\kappa_1 = 0\,,
}
\label{eq:condi2}
\end{equation}
where the FS equations in the null frame has been exploited. The equations
(\ref{eq:condi1}) are in fact the equations of motion of the particles 
governed by (\ref{eq:null-act}) whereas (\ref{eq:condi2}) are simple 
geometrical identities. The momentum acquires the form 
\begin{equation*}
\mathsf p = (\mathsf p _- ^{''} - \mathsf p_-\kappa_1 )e_+
+ \mathsf p_- e_- - \mathsf p _- ' e_1 + \mathsf p _2 e_2\,. 
\end{equation*}
We remark at this point that it is not necessary to know
the form that $\Lambda_{++}$ must hold. 

A straightforward computation in (\ref{eq:eom-n}) leads us to identify the
independent components of the momentum in terms of the worldline curvatures, 
$\mathsf p_- = (L - 2L^* \kappa_1 + L^{*\,''})/2$ and $\mathsf p_2= 2L^{*\,'}
\kappa_2 + L^* \kappa_2 '$.
Thus, we can write the linear momentum in the null FS frame
\begin{eqnarray}
\mathsf p &=&-\frac{1}{2}\left[ \left( L - 2 L^*\,\kappa_1 + L^{*\, ''} \right)\,
\kappa_1 - \left( L - 2L^{*}\,\kappa_1 + L^{*\,''}\right) ^{''}
\right] \,e_+ \nonumber \\
&+& \frac{1}{2} \left( L -2 L^{*}\,\kappa_1 + L^{*\,''}\right) \,e_- 
- \frac{1}{2} \left( L - 2L^{*}\kappa_1 +  L^{*\,''} \right)'\,e_1 
\nonumber \\
&+& \left( 2L^{*\,'}\,\kappa_2 + L^* \,\kappa_2 ' \right) \,e_2\,.
\label{eq:PP}
\end{eqnarray}
This is the general expression for the momentum associated to 
(\ref{eq:null-act}). It is worth pointing out that the momentum
(\ref{eq:PP}) is completely determined by two independent 
components $\mathsf p_-$ and $\mathsf p_2$ in the 4-dimensional case.
In fact, this is bequeathed from the theory of deformations where in 
order to preserve null curves in the variational procedure, two independent 
normal variations are necessary \cite{null}. 

To project the conservation law (\ref{eq:eom-n}) into the null FS 
frame is equivalent to express equations (\ref{eq:condi1}) and (\ref{eq:condi2}) 
in terms of the independent components of the momentum (\ref{eq:PP}). 
The first equation of (\ref{eq:condi2}) is
\begin{equation}
 \mathsf E_- = L' - L^*\,\kappa_1 ' = 0\,.
\end{equation}
This is merely an identity based in the chain rule from ordinary
calculus. The second equation of (\ref{eq:condi2}), $\mathsf E_1$, is just a null identity.
Now, the second equation of (\ref{eq:condi1}) is written as
\begin{equation}
 \mathsf E_2 = (L - 2L^*\,\kappa_1 + L^{*\,''})\kappa_2 +
2\left[ \left( L^{*\,2} \kappa_2 
\right)'/L^*\right]'  = 0\,.
\label{eq:eom-n2}
\end{equation}
This equation determines $\kappa_2$ in terms of $\kappa_1$.
Finally, the first equation of (\ref{eq:condi1}) results
\begin{eqnarray}
 \mathsf E_+ &=& \left( L - 2L^{*}\kappa_1 + L^{*\,''}\right)^{'''} -
2 \left( L - 2L^*\kappa_1 + L^{*\,''} \right)' \kappa_1 \nonumber \\
&-& \left( L - 2L^{*}\kappa_1 + L^{*\,''} \right) \kappa_1 ' + 
2\left[ \left( L^{*\,2} \kappa_2 \right)'/L^*\right]\kappa_2 =0\,. 
\label{eq:eom-n1}    
\end{eqnarray}
The expressions (\ref{eq:eom-n2}) and (\ref{eq:eom-n1}) determine the
equations of motion governing the dynamics of particles described by
the action (\ref{eq:null-act}) which do not appear to be tractable in
general. In fact, in the most simple cases they are two coupled 
differential equations whose solutions are null helices \cite{ferra1,ferra3}. 
There, the equations of motion can be integrated and expressed in terms 
of the mass and the spin of the particle \cite{null}. We must remark that
in a $2 + 1$ ambient Minkowski spacetime, besides $\kappa_2 = 0$,
the momentum component $\mathsf p_2$ disappears and the only equation of motion
is $\left( \mathsf p_- ^{''} - \mathsf p_- \kappa_1 \right)'
- \mathsf p_- ^{'} \kappa_1 = 0$. This latter equation also appears to be
intractable in general but surprisingly we find that for an arbitrary Lagrangian
$L$ it is possible to reduce the order of the equation of motion. We show
briefly how this comes about. In general, the equations of motion are equivalent
to the associated constants of motion given by the first and second Casimir 
invariants (See for example \cite{ferras} for a proof of this statement.).
By putting expressions (\ref{eq:M}) and (\ref{eq:S}) together we find that
\begin{equation}
L^{*\,'}\left( {\mathsf p}_- ^{2}\right)'  - L^* \left[ \left( 
{\mathsf p}_- ' \right)^{2} + M^2 \right]  + (L - L^{*\,''}){\mathsf p}_- ^2 
+ 2S\mathsf p_- = 0\,.
\label{eq:MSE}
\end{equation}
The immediate implication of this ODE in $\kappa_1$  is that it is equivalent
to the original equation of motion besides reduced in the order.
In a $3+1$ ambient spacetime, the integration for an arbitrary $L$ can be 
treated along the same lines but the computation is rather involved. 

We survey the application of the formalism by considering first a model 
for particles, in a $3+1$ ambient spacetime, given by a correction to 
the pseudo-arclenght parameter Lagrangian, $L = 2\left( \alpha + \beta\,
\kappa_1\right) $, where $\alpha$ and $\beta$ are constants. Obviously, 
$L^* = 2\beta$. The associated linear momentum is given by
\begin{equation} 
\mathsf p = -[\beta \kappa_1 ^{''} + (\alpha  - \beta \kappa_1)\kappa_1 ]
e_+ + (\alpha - \beta \kappa_1)e_- + \beta \kappa_1 '\,e_1
+ 2\beta \kappa_2 ' \, e_2 \,.
\end{equation}
Hence, from (\ref{eq:eom-n2}) and (\ref{eq:eom-n1}), the equations of 
motion are 
\begin{eqnarray}
\beta \kappa_1 ^{'''}
- \frac{3\beta}{2} \kappa_1 ^{' \,2} - \beta \kappa_2 ^{' \,2} + \alpha 
\kappa_1 ' &=& 0 \,, \\
2\beta \kappa_2 ^{''} - \beta \kappa_1 \kappa_2 + \alpha \kappa_2 &=&0\,,
\end{eqnarray}
which can be integrated and expressed in terms of the Casimir invariants. 
Recently, these equations of motion have been extensively studied in 
\cite{ferra1,ferra3}. 

Another example more complicated than the previous one can be 
found in a $2+1$ ambient spacetime. Consider $L= 2\left( \alpha + 
\beta\,\kappa_1 ^2 \right)$ with $\alpha$ and $\beta$ being constants. 
%We have, $L^* = 4 \beta\kappa_1, L^{*\,'}= 4\beta \kappa_1 ^{'}$ and $L^{*\,''}= 
%4\beta \kappa_1 ^{''}$. 
This model resembles to the $1+1$ timelike effective model for a relativistic
kink in the field of a soliton \cite{pichu}. The corresponding linear momentum is 
\begin{eqnarray}
\mathsf p &=& \left\lbrace \beta \left( -3 \kappa_1 ^2 + 2\kappa_1 '' \right)'' - \left[ \alpha + \beta \left( -3 \kappa_1 ^2 + 2\kappa_1 '' \right) 
\right] \kappa_1 \right\rbrace  e_+ \nonumber \\
&+&   \left[ \alpha + \beta \left(-3 \kappa_1 ^2 + 2\kappa_1 ''
\right) \right]  e_- - \beta \left( -3\kappa_1 ^2 + 2\kappa_1 '' 
\right) ' e_1 \,.
\end{eqnarray}
From (\ref{eq:eom-n1}) we obtain the equation of motion
\begin{equation}
\fl
\left\lbrace \beta \left( -3\kappa_1 ^2 + 2\kappa_1 '' \right)'' - 
2 \left[ \alpha + \beta \left( -3 \kappa_1 ^2 + 2\kappa_1 '' \right) 
\right]\kappa_1 \right\rbrace ' +\alpha \kappa_1 ' + \beta \left( -
3 \kappa_1 ^2 + 2\kappa_1 '' \right) \kappa_1 ' = 0,
\label{eq:eomk2}
\end{equation}
which can be integrated immediately to give a fourth-order ODE in $\kappa_1$
\begin{equation}
\kappa_1 ^{(4)} - 5\kappa_1 \kappa_1 '' - \frac{5}{2} (\kappa_1 ')^{2} +
\frac{5}{2} \kappa_1 ^3 - \gamma\,\kappa_1 - \lambda_{(3)} = 0\,,
\label{eq:reduced}
\end{equation}
where $\gamma= \alpha/2\beta$ and $\lambda_{(3)}$ is an integration 
constant which, in principle, can be written in terms of the Casimir 
invariants. One can go further on the integration of the Eq. (\ref{eq:eomk2})
if we appeal to the expression (\ref{eq:MSE}). The original equation 
of motion is equivalent to
\begin{equation}
\fl
\eqalign{
2\beta\kappa_1 \left\lbrace \left[ \alpha + \beta \left( -3 \kappa_1 ^2 + 
2\kappa_1 '' \right) \right]^{'\,2} + M^2 \right\rbrace  -
2\beta \kappa_1 ' \left\lbrace \left[ \alpha + \beta \left( -3 \kappa_1 ^2 + 
2\kappa_1 '' \right) \right]^{2}\right\rbrace ' 
\\
- \left( \alpha + \beta \kappa_1 ^2 - 2\beta \kappa_1 '' \right)
\left[ \alpha + \beta \left( -3 \kappa_1 ^2 + 2\kappa_1 '' \right) 
\right]^{2} + S \left[ \alpha + \beta \left( -3 \kappa_1 ^2 + 2\kappa_1 '' 
\right) \right]= 0\,,}
\label{eq:int}
\end{equation}
where $M^2$ is the first Casimir invariant given by
\begin{equation}
\eqalign{
M ^2 = \left\lbrace \left[ \alpha + \beta\left( \kappa_1 ^2 - 
2\beta \kappa_1 ''\right) \right]^2 \right\rbrace '' - 3\beta ^{2}
\left[ \left( -3 \kappa_1 ^2 + 2\kappa_1 '' \right) ' \right] ^2  \\
-2 \left[ \alpha + \beta\left( \kappa_1 ^2 - 
2\beta \kappa_1 ''\right) \right]^2 \kappa_1 \,,}
\end{equation}
and $S$ is the associated second Casimir invariant which becomes
\begin{equation}
\fl
\eqalign{
 S = - \left[ \alpha + \beta \left( \kappa_1 ^2 -  2\kappa_1 '' \right) \right]
\left[ \alpha + \beta \left( -3 \kappa_1 ^2 + 2\kappa_1 '' \right) \right] - 
4\beta ^{2} \kappa_1 ' \left( -3 \kappa_1 ^2 + 2\kappa_1 '' \right)'  \\
+ 4 \beta \kappa_1 \left[ -\alpha \kappa_1 + \beta \left( -3 \kappa_1 ^2 
+ 2\kappa_1 '' \right) '' - \beta \kappa_1 \left( - 3 \kappa_1 ^2 + 2\kappa_1 '' 
\right)  \right].}
\end{equation}
Despite we have enormously reduced the order of the original equation of motion,
the equivalent equation (\ref{eq:int}) turns out to be complicated as opposed 
to the equation (\ref{eq:reduced}). The main benefit of (\ref{eq:reduced}) resides
in its simplicity. In fact, alike to (\ref{eq:reduced}) is the resulting equation 
of motion for a particle described by a model linear in 
$\kappa_1$ in a $3+1$ ambient spacetime \cite{ferra3}.

\section{Concluding remarks}

In this paper we have analysed worldline theories by obtaining the 
associated conserved linear momentum. This has been done by means of 
an auxiliary variables method. The main advantage of the method is based in the
reduction of the higher order derivative nature of the fields, obtaining
considerable simplication in the variational procedure and avoiding awkward 
computations. We have tailored this auxiliary 
variables method to the FS frame of each curve, either timelike of lightlike.
Based on Poincar\'e and reparametrization 
invariance of the action the conservation of the momentum leads us 
to the full mechanical content of the worldline theories. Equations 
(\ref{eq:eom}) and (\ref{eq:eom-n}) provide the dynamics 
for arbitrary Lagrangians $L(k_1)$ and $L(\kappa_1)$, when they are
implemented by the FS frame projections. We showed that the auxiliary 
variables method is in fact a powerful alternative to study embedded theories. 
Although originally it was implemented to study general surfaces characterized by
their extrinsic geometry, like the lipid membranes \cite{jemal}, 
its application is immediate to relativistic brane models under interaction
with other fields.
The complete integrability of the equations of motion faces several technical 
difficulties when one tries to integrate them for a general Lagrangian $L(\kappa_1)$. It
seems to be intractable in general. We explored some examples to see 
our machinery at work.
For the simplest cases, like a constant and linear in $\kappa_1$, the 
integrability is a well known fact. Nevertheless, we have shown the existence 
of other model, quadratic in $\kappa_1$, in a $2+1$ ambient spacetime where we 
have also obtained integrability. Work along this issue is in progress.

\ack 
E.R. cordially would like to thank Alberto Molgado for useful comments and 
suggestions. Also, E. R. has been benefited from conversations with Jemal Guven 
and Eloy Ay\'on. E. R. acknowledges partial support from CONACyT 
under grant 51111. R.C. acknowledges partial support by grants COFAA, EDI, 
SIP-20071135 and CONACyT J1-60621-I. N.B. acknowledges partial support by grant 
SEP-2003-C02-43780. The authors would like to thank 
anonymous referees who suggested several important improvements.

\appendix

\section{}

\subsection{Arclength infinitesimal geometry}
\label{sec:arc}

In this Appendix we express the main variations necessary to
develop the dynamics from a general action functional using 
the proper time as the parameter to describe the curve 
\cite{ACG,nesterenko,kalman}. 

For an infinitesimal deformation of a timelike worldline, 
$X^\mu (\xi) \to X^\mu (\xi) + \delta X^\mu (\xi)$, we 
can decompose the deformation with respect to the FS basis as
\begin{equation}
\delta X = \Phi\,X' + \Psi_i\,\eta_i\,.
\end{equation}
The tangential projection can be identified with 
reparametrizations of the worldline. The tangential 
deformation of the proper time (\ref{eq:1})  is, 
$\delta_\Vert d \tau = \Phi'\,d\tau$.
It follows straightforwardly that
\begin{eqnarray}
\delta d \tau &= \left( \Phi' + k_1 \Psi_1 \right)\,d\tau =
-(X'\cdot  \frac{d}{d\tau}\delta X)\,d\tau \,,\\
\left[ \delta , \frac{d}{d\tau}\right]&= -\left( 
 \Phi' + k_1 \Psi_1 
\right)\,\frac{d}{d\tau}= \left( X'\cdot \frac{d}{d\tau} \delta X 
\right) \frac{d}{d\tau}\,,  
\label{eq:T1}
\end{eqnarray}
which help us to recognize the dependence on the parametrization
of the proper time and the FS basis through the prime derivatives
and show that the operations $\delta$ and $d/d\tau$ do not
commute \cite{ACG}.

\subsection{Pseudo arclength infinitesimal geometry}
\label{sec:p-arc}

For an infinitesimal deformation of a null curve, $\x^\mu (\rho)
\to \x^\mu (\rho) + \delta \x^\mu (\rho)$, we can express the
deformation with respect to the null FS frame as
\begin{equation}
 \delta \x = \epsilon_+ e_+ + \epsilon_-e_- + \epsilon_1 e_1 
+ \epsilon_2 e_2\,.
\end{equation}
Similarly as in the timelike case, the tangential projection can 
be identified with reparametrizations of the null curve such that
the tangential infinitesimal deformation of the pseudo-arclength
(\ref{eq:pseudo}) is given by $\delta_\Vert d\sigma = \epsilon_+ '
d\sigma$. 

To preserve the null character of the curve, the condition 
$\delta(\dot{\x}\cdot \dot{\x})=0$, leads us to the constraint 
\begin{equation}
\epsilon_1 + \epsilon_- ' = 0\,,
\label{eq:cond}
\end{equation} 
on the components of the deformation. Explicitly, this condition
is equivalent to $e_+ \cdot \frac{d}{d\sigma}\delta \x =0$.

In the null FS frame one can show analogously the identities
\begin{equation}
\eqalign{
\delta d\sigma = \frac{1}{2}\left( 2\epsilon_+ ' - \epsilon_- '''
+ \kappa_1 ' \epsilon_- + \kappa_2 \epsilon_2 \right) d\sigma \nonumber
= -\frac{1}{2}\left( e_1 \cdot \frac{d^2}{d\sigma^2}\delta \x\right)
d\sigma \,,
\\
\left[\delta , \frac{d}{d\sigma} \right]  = -\frac{1}{2}\left( 
2\epsilon_+ ' - \epsilon_- ''' + \kappa_1 ' \epsilon_- + \kappa_2 
\epsilon_2 \right)\frac{d}{d\sigma}\nonumber
= \frac{1}{2}\left( e_1 \cdot \frac{d^2}{d\sigma^2}\delta \x \right)
\frac{d}{d\sigma}
\,.}
\label{eq:TT2}
\end{equation}
Similarly, as in the timelike case, the operations $\delta$ and 
$d/d\sigma$ do not commute \cite{null}.

\section{Noether invariants in the lightlike case}

\label{sec:noether}

The writing of the first variation of the action (\ref{eq:null-act}) 
in the form
\begin{equation*}
\delta S = \int d\sigma \, \mathsf E \cdot \delta \mathsf X + \int d\sigma\, {\mathsf Q}'\,,
\end{equation*}
where $\mathsf E^\mu$ stands for the EL derivative associated to $\mathsf X$, entails
the identification of the associated Noether charge $\mathsf Q$ given by \cite{ACG}
\begin{eqnarray}
\fl
 \mathsf Q = \mathsf p \cdot \delta \mathsf X - \frac{1}{2}\left[ \left( L - L^{*\,''} + L^* \,
\kappa_1 \right)e_1 + 2L^{*\,'} e_- + 2 L^* \kappa_2 \,e_2  \right] \cdot \frac{d}{d\sigma}\delta
\mathsf X \nonumber \\
+ \frac{1}{2}\left[ \left( 2 L^{*}\kappa_1  - L^{*\,'} \right)e_1  + 3 L^*  \,e_- \right] \cdot 
\frac{d^2}{d\sigma^2}\delta \mathsf X + \frac{1}{2}L^* \, e_1 \cdot \frac{d^3}{d\sigma^3}
\delta \mathsf X  \,.
\end{eqnarray}
It is clear that if the deformation $\delta \x ^\mu$, is a constant infinitesimal
deformation, $\delta \x ^\mu = \varepsilon^\mu$, and assuming $\mathsf Q = 
\varepsilon^\mu \mathsf p_\mu$, we are able to recuperate the expression for the linear 
momentum (\ref{eq:PP}). The first Casimir invariant of the Poincar\'e group, 
$M^2 = \mathsf p ^2$, results
\begin{eqnarray}
M^2 &=&
(\mathsf p_- ^2)^{''} - 3 (\mathsf p_- ')^2 - 2 \mathsf p_- ^2 \,\kappa_1
- \mathsf p_2 ^2 \nonumber \\
&=& \frac{1}{4}\left[ \left( L - 2L^* \kappa_1 + L^{*\,''}
\right)^2 \right] ^{''} - \frac{3}{4} \left[ \left( L - 2L^* \kappa_1 
+ L^{*\,''} \right)'\right]^2 
\nonumber \\
&-& \frac{1}{2}\left( L - 2L^* \kappa_1 + L^{*\,''} \right)^2 \kappa_1 -  
\left[ \left( L^{*\,2} \kappa_2 
\right)'/L^*\right]^2 \,.
\label{eq:M}
\end{eqnarray}

On the other hand, by specializing the deformation  $\delta \x ^\mu$
to Lorentz transformations, $\delta \x ^\mu = \omega^\mu {}_\nu \x^\nu$ 
with $\omega_{\mu \nu} = - \omega_{\nu \mu}$ and assuming $\mathsf 
Q = \omega_{\mu \nu} \mathsf M^{\mu \nu}$, we obtain the conserved angular 
momentum 
\begin{equation}
 \mathsf M ^{\mu \nu} = \mathsf p^{[\mu}X^{\nu]} + \frac{1}{2}\left( L - L^{*\,''} \right)
e_+ ^{[\mu} e_1 ^{\nu ]} + L^{*\,'} e_+ ^{[\mu} e_- ^{\nu ]} + 
L^* \kappa_2 e_+ ^{[\mu} e_2 ^{\nu ]} + L^{*} e_- ^{[\mu} e_1 ^{\nu ]}.
\end{equation}
If the particle moves in a $3+1$ ambient spacetime, to extract the spin content of 
the particle models governed by the action 
(\ref{eq:null-act}), we introduce the Pauli-Lubanski pseudo-vector, $S_\mu = \frac{1}{2\sqrt{|M^2|}} 
\varepsilon_{\mu \nu \rho \sigma} {\mathsf p}^\nu \mathsf 
M^{\rho \sigma}$, which results
\begin{equation}
\fl
\eqalign{
S_\mu 
= \frac{1}{2\sqrt{|M^2|}}\left\lbrace - \left[ \mathsf p_- ' L^* 
\kappa_2 + \frac{1}{2} \mathsf p_2 (L - L^{*\,''}) \right]
e_{+\,\mu}  + \left( \mathsf p_2 L^*\right) e_{-\,\mu}  \right. 
\\ 
\left.
- \left( \mathsf p_2 L^* \right)'\,e_{1\,\mu} - \left[ \left( 
\mathsf p_- \,\kappa_1 - \mathsf p_- ^{''} \right) L^* + \frac{1}{2}
\mathsf p_- \left( L - L^{*\,''} \right) + \mathsf p_- '\,L^{*\,'}
\right] e_{2\,\mu} \right\rbrace ,
}
\end{equation}
where $\varepsilon_{\alpha \beta \rho \sigma}$ is the Levi-Civita 
tensor density and we have used the following convention 
$\varepsilon_{\alpha \beta \rho \sigma} e_+ ^\alpha e_- ^\beta 
e_1 ^\rho e_2 ^\sigma = +1$. The second Casimir invariant of the 
Poincar\'e group is 
\begin{eqnarray}
\fl
4|M^2|S^2 = - \left[ \left( \mathsf p_-\,\kappa_1 - 
\mathsf p_- ^{''} \right) \,L^* + \frac{1}{2}\mathsf p_- 
\left( L - L^{*\,''} \right) + \mathsf p_- '\,L^{*\,'} \right]^2
- \left[ (\mathsf p_2 L^*)' \right] ^2
\nonumber \\
- 2 \left[ \mathsf p_-' L^* \,\kappa_2 + \frac{1}{2}\mathsf p_2 
\left( L - L^{*\,''} \right) \right] (\mathsf p_2 L^*) \,.
\end{eqnarray}

Now, for a $2+1$ ambient spacetime, the spin content of particles can be extracted from
the spin pseudo-vector, $ \mathsf J_\mu = \varepsilon_{\mu \alpha \beta}M^{\alpha \beta}$, 
resulting in
\begin{eqnarray}
 \mathsf J_\mu 
&=& \varepsilon_{\mu \alpha \beta}\mathsf p^\alpha \x^{\beta} - \frac{1}{2} \left( L -
L^{*\,''}\right) e_{+\,\mu} - L^{*\,'}e_{1\,\mu} + L^{*} e_{-\,\mu}\,.
\end{eqnarray}
Thus, we have the second Casimir, $S=\mathsf J \cdot \mathsf p$, given by
\begin{equation}
 S = - \frac{1}{2} \left( L -
L^{*\,''}\right)\mathsf p_- - L^{*\,'}\mathsf p_- ^{'} + L^{*}
\left( \mathsf p_- ^{''} - \mathsf p_- \kappa_1 \right)\,. 
\label{eq:S}
\end{equation}
It should be clear at this point that the Noether invariants of the underlying 
Poincar\'e symmetry are expressed in terms of the geometry of the worldline. In addition,
the spin content of this sort of particles depends heavily of the worldline curvatures.

\end{document}